\documentstyle[12pt,rsfs]{article}

\title{On the numerical integration of motion for rigid
       polyatomics: The modified quaternion approach \\ [6pt]}

\author{\sc Igor~P.~Omelyan \\ [1.5ex]
{\small \em Institute for Condensed Matter Physics,
            National Ukrainian Academy of Sciences,} \\ [-8pt]
{\small \em 1~Svientsitsky St., UA-290011 Lviv, Ukraine.
            E-mail: nep@icmp.lviv.ua} \\
\date{}}

\newcommand{\bms}[1]{\mbox{\boldmath $#1$}}
\newcommand{\bvs}[1]{\mbox{\scriptsize\boldmath $#1$}}

\begin{document}

\setlength{\abovedisplayskip}{16pt plus4pt minus4pt}
\setlength{\belowdisplayskip}{\abovedisplayskip}
\setlength{\abovedisplayshortskip}{10pt plus2pt minus2pt}
\setlength{\belowdisplayshortskip}{\abovedisplayshortskip}

\maketitle

\vspace{1cm}

\begin{abstract}

A revised version of the quaternion approach for numerical integration
of the equations of motion for rigid polyatomic molecules is proposed.
The modified approach is based on a formulation of the quaternion dynamics
with constraints. This allows to resolve the rigidity problem rigorously
using constraint forces. It is shown that the procedure for preservation
of molecular rigidity can be realized particularly simply within the
Verlet algorithm in velocity form. We demonstrate that the presented
method leads to an improved numerical stability with respect to the
usual quaternion rescaling scheme and it is roughly as good as the
cumbersome atomic-constraint technique.

\end{abstract}

\newpage

\section{Introduction}

\hspace{1em}  Many models of statistical mechanics deal with systems
composed of classical rigid molecules. The method of molecular dynamics
(MD) is widely applied for studying such systems. All known MD techniques
appropriate to simulate molecular liquids can be split into three main
approaches. In the first two approaches, time evolution of the system
is considered in view of translational and rotational motions. These
approaches differ between themselves by parameters which are used to
represent the rotational degrees of freedom. In the classical scheme
[1, 2], an orientation of the molecule is defined in terms of three
Eulerian angles. As is well known [3], the equations of motion are
singular in this case. To avoid the singularities, Barojas {\em et al}
[4, 5] have used two different sets of the Eulerian angles, each of which
is applied in dependence on the orientation of the molecule. However, this
procedure involves additional complex transformations with transcendental
functions.

In the second approach the rotation motion of a molecule is described
without involving Eulerian angles. Cheung [6] has shown how to remove
the singularities using special properties of diatomic molecules. For
these molecules, Singer [7] has derived rotational equations of motion
in terms of radius-vectors passed from one atom to another within the
same molecule. An extension of this scheme to triatomic molecules was
considered also [8, 9]. An alternative scheme has been proposed by Evans
{\em et al} [10--12], where using so-called quaternions [3, 13, 14] leads
to a singularity free algorithm for rigid polyatomics.

In the third approach, proposed originally by Ryckaert {\em et al} [15],
the cartesian equations of motion are applied with respect to individual
atoms. The total force on a particle appears as the sum of the force
deriving from the potential energy and the force arising due to holonomic
constraints. These atomic constraints must be in part rigid bonds in part
linear relations to provide the rigidness of arbitrary polyatomics [16].

Apart from removing singularities a benefit derived from the last two
approaches lies in the avoidance of time-consuming trigonometric functions.
However, on integrating the equations of motion numerically, one additional
difficulty appears here, namely, the problem of exact conservation of
molecular rigidity. In the usual integration algorithms the rigidity can
not be conserved with the precision better than that of evaluating the
atom trajectories. For overcoming this drawback, it is necessary either to
perform the rescaling of quaternions [11] or, within the atomic-constraint
technique, to find solutions for a complete system of nonlinear equations
[15, 16] at each step of the integration.

From the aforesaid, a natural question appears about the existence of a
scheme which is free of all these drawbacks and yet has all advantages
inherent in the mentioned above approaches. In the present paper we develop
the idea of using quaternions to treat rotational motion. Section~2 is
devoted to a general formulation of the quaternion dynamics with constraints.
Applications of this approach to particular algorithms are considered in
Sec.~3. The problem of how to adapt the Verlet algorithm to integrate the
quaternion equations of motion is also solved there. In Sec.~4 various
approaches are compared and discussed. Some concluding remarks are
given in Sec.~5.

\vspace{12pt}

\section{Quaternion dynamics with constraints}

\hspace{1em}  We consider a system of $N$ identical rigid molecules with
mass $m$, composed of $M$ point atoms. In the molecular approach, evolution
of the system in time is separated into translational and rotational
motions. The translational motion is applied to the molecule as a whole
and can be described by the $3N$ ($i=1,\ldots,N$) Newton equations
$m \frac{{\rm d}^2 \bms{r}_i}{{\rm d} t^2} = \sum_{j;a,b}^{N;M}
\bms{F}_{ij}^{ab} (|\bms{r}_i^a-\bms{r}_j^b|)$, where $\bms{r}_i$ and
$\bms{r}_i^a$ are the positions of the center of mass and atom $a$ of
molecule $i$ in the laboratory fixed coordinate system L, respectively,
and $\bms{F}_{ij}^{ab}$ are the atom-atom forces between two different
molecules.

In order to analyze rotational motions, we introduce the sets $\bms{e}
\equiv (\bms{e}_1, \bms{e}_2, \bms{e}_3)$ and $(\bms{u}_1^i,\bms{u}_2^i,
\bms{u}_3^i)=\bms{\rm A}_i \bms{e}$ of orthogonal unit vectors
characterizing the L-system and the moving coordinate system S$^i$
attached to molecule $i$, respectively, where $\bms{\rm A}_i$ is a
rotational matrix. The angular velocity $\bms{\mit \Omega}_i$ of the
$i$-th molecule is defined as ${\rm d} \bms{u}_\alpha^i /{\rm d} t =
\bms{\mit \Omega}_i \bms{\times} \bms{u}_\alpha^i$. The principal
components, ${\mit \Omega}_1^i \bms{u}_1^i+{\mit \Omega}_2^i \bms{u}_2^i+
{\mit \Omega}_3^i \bms{u}_3^i = \bms{\mit \Omega}_i$, of angular velocities
($i=1,\ldots,N$) obey the $3N$ Euler equations [1]:
\begin{equation}
J_\alpha \frac{{\rm d} {\mit \Omega}_\alpha^i}{{\rm d} t} =
K_\alpha^i(t) + \Big(J_\beta-J_\gamma\Big) {\mit \Omega}_\beta^i(t)
{\mit \Omega}_\gamma^i(t) \ , \ \ \ \ \ \ \
\end{equation}
where $(\alpha,\beta,\gamma)=(1,2,3)$; $(2,3,1)$ and $(3,1,2)$. Here
$J_1$, $J_2$ and $J_3$ are the moments of inertia along principal axes
of the molecule, $\sum_{j;a,b}^{N;M} \bms{\delta}_i^a \bms{\times}
\bms{F}_{ij}^{ab} = k_1^i \bms{e}_1+k_2^i \bms{e}_2+k_3^i \bms{e}_3 =
K_1^i \bms{u}_1^i+K_2^i \bms{u}_2^i+K_3^i \bms{u}_3^i$ is the torque
exerted on molecule $i$ with respect to its center of mass due to
the interactions with the other molecules, $\bms{K}_i=\bms{\rm A}_i
\bms{k}_i$ and $\bms{\delta}_i^a=\bms{r}_i^a-\bms{r}_i$. Let $\bms
{\Delta}^a=(\Delta^a_1, \Delta^a_2, \Delta^a_3)$ be a vector-column of
positions for atom $a$ within the molecule in the S$^i$-system, i.e.,
$\bms{\delta}_i^a=\Delta^a_1 \bms{u}_1^i+\Delta^a_2 \bms{u}_2^i+
\Delta^a_3 \bms{u}_3^i$. Then the positions of atoms in the L-system
at time $t$ are $\bms{r}_i^a(t) = \bms{r}_i(t) + \bms{\rm A}_i^{\bvs{+}}(t)
\bms{\Delta}^a$, where $\bms{\rm A}^{\bvs{+}}$ denotes the matrix transposed
to $\bms{\rm A}$.

It is a common practice to define an orientation of the S$^i$-system with
respect to the laboratory frame in terms of three Eulerian angles. A
numerical integration of the corresponding equations of motion has been
performed in early investigations [1, 2]. As was soon realized, however,
this procedure is very inefficient because of the singularities whenever
the azimuthal angle of a molecule takes the value $0$ or $\pi$ [11]. It
has been shown in later investigations [10, 11] that at least four
orientational parameters per molecule (quaternion) must be used to
avoid the singularities.

The orientational matrix $\bms{\rm A}_i = \bms{\rm A} (\bms{q}_i)$ in
terms of the quaternion $\bms{q}_i \equiv (\xi_i,\eta_i,\zeta_i,\chi_i)$
is given by [10, 11]:
\begin{equation}
\bms{\rm A} (\bms{q}_i) =
\left( \begin{array}{ccc}
-\xi_i^2+\eta_i^2-\zeta_i^2+\chi_i^2 & 2(\zeta_i \chi_i - \xi_i \eta_i) &
2 (\eta_i \zeta_i + \xi_i \chi_i) \\
-2(\xi_i \eta_i + \zeta_i \chi_i) & \xi_i^2-\eta_i^2-\zeta_i^2+\chi_i^2 &
2(\eta_i \chi_i - \xi_i \zeta_i) \\
2(\eta_i \zeta_i - \xi_i \chi_i) & -2(\xi_i \zeta_i + \eta_i \chi_i) &
-\xi_i^2-\eta_i^2+\zeta_i^2+\chi_i^2
\end{array}
\right) \ \ \ \
\end{equation}
and the time derivative of the quaternion is expressed via principal
components of angular velocity as follows
\begin{equation}
\bms{\dot q}_i \equiv
\left( \begin{array}{c}
\dot \xi_i \\ \dot \eta_i \\ \dot \zeta_i \\ \dot \chi_i
\end{array} \right)
= \displaystyle \frac12 \left( \begin{array}{cccc}
-\zeta_i & -\chi_i & \eta_i & \xi_i \\
\chi_i & -\zeta_i & -\xi_i & \eta_i \\
\xi_i & \eta_i & \chi_i & \zeta_i \\
-\eta_i & \xi_i & -\zeta_i & \chi_i
\end{array}
\right)
\left( \begin{array}{c}
{\mit \Omega}_1^i \\ {\mit \Omega}_2^i \\ {\mit \Omega}_3^i \\ 0
\end{array} \right)
\equiv \frac12 \bms{\rm Q}(\bms{q}_i) \bms{\Omega}_i \ , \ \ \ \
\end{equation}
where the matrix $\bms{\rm Q}$ is the function of $\bms{q}_i$. It is worth
to underline that the matrix $\bms{\rm A}$ is a rotational one if the
quaternion satisfies the equality $\bms{q}_i^2=\xi_i^2+\eta_i^2+\zeta_i^2+
\chi_i^2=1$. Differentiating the relation (3) over time yields
\begin{equation}
\bms{\ddot q}_i=\frac12 \left( \bms{\rm Q}(\bms{\dot q}_i)
\bms{\Omega}_i + \bms{\rm Q}(\bms{q}_i) \bms{\dot{\Omega}}_i
\right) \, , \ \ \ \ \ \
\end{equation}
where $\bms{\Omega}_i = 2 \bms{\rm Q}^{-1}(\bms{q}_i) \bms{\dot q}_i$.
It is trivial to find that the inverse matrix $\bms{\rm Q}^{-1}=
\bms{\rm Q}^{\bvs{+}}$ since the $4 \times 4$ matrix $\bms{\rm Q}$
is orthogonal when $\bms{q}_i^2=1$. We have augmented the angular
velocity vector to involve square matrices using the result $\bms{q}_i
\bms{\cdot} \, \bms{\dot{q}}_i=\xi_i \dot \xi_i+\eta_i \dot \eta_i+
\zeta_i \dot \zeta_i+\chi_i \dot \chi_i=0$ which follows from the
equality $\bms{q}_i^2=1$.

Then, using the Newton and Euler equations (1), we obtain the coupled set
of $7N$ second-order differential equations of motion ${\scr F}(\{ \bms{r}_i,
\bms{\ddot r}_i, \bms{q}_i, \bms{\dot q}_i, \bms{\ddot q}_i\})=0$ in terms
of the $7N$ generalized coordinates $\{\bms{r}_i, \bms{q}_i\}$. If an
initial state $\{\bms{r}_i(t_0), \bms{\dot r}_i(t_0), \bms{q}_i(t_0),
\bms{\dot q}_i(t_0)\}$ is specified, the evolution $\{\bms{r}_i(t),
\bms{q}_i(t)\}$ of the system can be unambiguously determined.

Let us look for an analytical solution of the quaternion equations of
motion by writing
\begin{equation}
\bms{q}_i(t)=\sum_{p=0}^P \bms{q}_i^{(p)}(t_0) \frac{(t-t_0)^p}{p!} \ ,
\ \ \ \ \
\end{equation}
where $\bms{q}_i^{(p)}(t_0)$ denotes the $p\,$-fold time derivative of
$\bms{q}_i$ at time $t_0$. It is easy to check from the structure of
equation (3) that arbitrary-order time derivatives of the quaternion
constraint $\sigma_i(t) \equiv \bms{q}_i^2(t)-1=0$ are equal to zero,
i.e., $\bms{q}_i \bms{\cdot} \bms{\dot q}_i=0$, $\bms{\dot q}_i^2+
\bms{q}_i \bms{\cdot} \bms{\ddot q}_i=0$ and so on. Therefore, if all
terms $(P \to \infty)$ of the Taylor's expansion (5) are taken into
account and initially all the constraints are satisfied, $\sigma_i(t_0)=0$,
they will be fulfilled at later times as well. In practice, however, the
equations of motion are not solved exactly, so that these constraints
will only be satisfied approximately. Let the integration algorithm used
involves an error in the coordinates of order ${\mit\Delta}t^{P+1}$, where
${\mit\Delta}t=t-t_0$ is the time step. In the simplest case of Taylor's
expansion (5), this is the order of the first omitted term. Then the same
order of uncertainties will be accumulated at each time step in conservation
of the molecular rigidity, i.e., $\sigma_i(t)={\cal O}({\mit\Delta}t^{P+1})$.
In such a case, molecules are collapsed or even destroyed in time. In the
usual version [11] of the quaternion method, to achieve the required rigidity
at all times, it was proposed to multiply each quaternion component,
associated with the same molecule, on the common factor $1 \Big/
\sqrt{\bms{q}_i^2}$ at every time step of the numerical integration
(the so-called rescaling scheme).

We consider now the question how to replace the crude quaternion
renormalization by a more natural procedure in the framework of a
systematic approach. The fact that quaternion components are not
independent, requires, in general, the necessity of introducing additional
forces, namely, $\bms{f}_i(t)=-\lambda_i(t) \bms{\nabla}_{\bvs{q}_i}
\sigma_i(t)=-2 \lambda_i(t) \bms{q}_i(t)$, which appear as a result of the
constraints. These virtual quaternion-constraint forces should be added to
the equations of motion (4) and, as a consequence, they modify the solution
(5) as follows
\begin{equation}
\bms{q}_i(t) =
\sum_{p=0}^P \bms{q}_i^{(p)}(t_0) \frac{(t-t_0)^p}{p!} +
\sum_{p=2}^P \bms{f}_i^{(p-2)}(t_0) \frac{(t-t_0)^p}{p!} \ ,
\ \ \ \ \ \ \\ [4pt]
\end{equation}
where $\bms{f}_i^{(p-2)}(t_0) = -2 \sum_{k=0}^{p-2} C_{p-2}^k \lambda_i^{(k)}
(t_0) \bms{q}_i^{(p-2-k)}(t_0)$ denote the $(p-2)$-fold time derivatives of
constraint forces, $\lambda_i^{(0)}(t_0)$ is a value of the Lagrange
multiplier $\lambda_i(t)$, and $\lambda_i^{(k)}(t_0)$ are its $k$-fold time
derivatives ($k=1,\ldots,P-2$) at time $t_0$. Differentiating (6), we
obtain $l$-fold time derivatives ($l=0,\ldots,P-2$) of $\bms{q}_i$ at
time $t$:
\begin{equation}
\bms{q}_i^{(l)}(t) =
\sum_{p=l}^P \bms{q}_i^{(p)}(t_0) \frac{(t-t_0)^{p-l}}{(p-l)!} -2
\sum_{p=\max\{2,l\}}^P \sum_{k=0}^{p-2} C_{p-2}^k
\lambda_i^{(k)}(t_0) \bms{q}_i^{(p-2-k)}(t_0)
\frac{(t-t_0)^{p-l}}{(p-l)!} \ . \
\end{equation}
In order to computer $P-1$ unknowns $\lambda_i^{(k)}(t_0)$, we have
merely to exploit the information contained in the constraint $\sigma_i(t)
\equiv \bms{q}_i(t)^2-1=0$. As this holds at any time, at least the first
$P-2$ time derivatives of $\sigma_i(t)$ must vanish. Then one obtains
$(p=1,...,P-2)$:
\vspace{1mm}
\begin{equation}
\sigma_i^{(p)}({q}_i(t)) \equiv \frac
{{\rm d}^p}{{\rm d} t^p} \sigma_i(t)
=2 \sum_{k=0}^{P-1} C_{p-1}^k \bms{q}_i^{(k)}(t)
\bms{q}_i^{(p-k)}(t) = 0 \ . \ \ \ \ \ \\ [1mm]
\end{equation}

In view of explicit expressions (7), the conditions (8) together with
the basic constraints $\bms{q}_i^2(t)=1$ constitute a system of $P-1$
nonlinear equations per molecule with respect to the same number of
unknowns $\lambda_i^{(k)}(t_0)$. The equations can be linearized and solved
in a quite efficient way by iteration. This is justified for ${\mit\Delta}t
\to 0$ because then the terms nonlinear in $\lambda_i^{(k)}(t_0)$ are
small. Thus the iteration procedure can be initiated by substituting
\mbox{$\lambda_i^{(k)}(t_0)=0$} in all nonlinear terms, and iterations
always converge rapidly to the physical solutions $\lambda_i^{(k)}(t_0)
\sim {\mit\Delta}t^{P-k-1}$. The contributions of quaternion-constraint
forces into the quaternion dynamics (6) are of order ${\mit\Delta}t^{P+1}$,
i.e., the same order as uncertainties of the integration algorithm (5), but
the rigidity is now fulfilled perfectly for arbitrary times in future. It
is worth emphasizing that these forces are imaginary and depend on details
of the numerical integration in a characteristic way, contrary to the real
bond forces in the atomic-constraint dynamics [15, 16]. They vanish if the
equations of rotational motion are solved exactly.

\vspace{12pt}

\section{Applying actual algorithms}

\subsection{Integration within the Gear method}

\hspace{1em}  Usually, the Gear predictor-corrector algorithm [17, 18] is
applied to integrate the equations of rotational motion [1, 11, 12]. In
particular, it has been used [11, 12] for the integration of the quaternion
equations. Within the Gear method the quaternions and their time derivatives
are predicted using the Pascal triangle $\bms{q}_i^{(l)}(t+{\mit\Delta}t) =
\sum_{p=l}^P \bms{q}_i^{(p)}(t) \frac{{\mit\Delta}t^{p-l}}{(p-l)!}$, where
$l=0,\ldots,P$ and $P$ is the order of the algorithm. Further, they are
corrected one or more times, using new values of torques as well as
rotational velocities and their time derivatives which are predicted and
corrected simultaneously with quaternion variables.

The Gear method can be modified within the quaternion-constraint dynamics
as follows. To simplify notations, we choose the fourth order scheme ($P=4$)
(the extension to arbitrary orders is trivial). Let $\bms{q}_i^{(l)}(t+{\mit
\Delta}t)$ (as well as $\bms{q}_i^{(l)}(t)$) be already defined quantities
after the last step of the corrector procedure. Then, according to the
constraint formalism, the variables $\bms{q}_i(t+{\mit\Delta}t)$,
$\bms{\dot q}_i(t+{\mit\Delta}t)$ and $\bms{\ddot q}_i(t+{\mit\Delta}t)$
$(l=0,1,2)$ transform into
\begin{eqnarray}
&&\bms{q'}_{\!i}(t+{\mit\Delta}t) = \bms{q}_i(t+{\mit\Delta}t) +
\bms{f}_i(t) {\mit\Delta}t^2/2 + \bms{\dot f}_i(t) {\mit\Delta}t^3/6 +
\bms{\ddot f}_i(t) {\mit\Delta}t^4/24 \ , \ \ \ \ \ \ \ \ \nonumber \\
&&\bms{\dot{q}'}_{\!i}(t+{\mit\Delta}t) = \bms{\dot q}_i(t+{\mit\Delta}t) +
\bms{f}_i(t) {\mit\Delta}t + \bms{\dot f}_i(t) {\mit\Delta}t^2/2 +
\bms{\ddot f}_i(t) {\mit\Delta}t^3/6 \ , \\
&&\bms{\ddot q'}_{\!i}(t+{\mit\Delta}t) =
\bms{\ddot q}_i(t+{\mit\Delta}t) + \bms{f}_i(t) +
\bms{\dot f}_i(t) {\mit\Delta}t + \bms{\ddot f}_i(t) {\mit\Delta}t^2/2 \ ,
\nonumber
\end{eqnarray}
where $\bms{f}_i(t) = -2 \lambda_i \bms{q}_i(t)$, $\bms{\dot f}_i(t) =
-2 (\lambda_i \bms{\dot q}_i(t) + \dot{\lambda}_i \bms{q}_i(t))$,
$\bms{\ddot f}_i(t) = -2 (\lambda_i \bms{\ddot q}_i(t) + 2 \dot{\lambda}_i
\bms{\dot q}_i(t) + \ddot{\lambda}_i \bms{q}_i(t))$ and $\lambda_i$,
$\dot \lambda_i$, $\ddot \lambda_i$ are values of the Lagrange multiplier
and its first and second time derivatives at time $t$.
The expressions (9) present, in fact, (in somewhat other notations) a
particular case ($P=4$) of generalized equations (7). Therefore,
the three unknowns $\lambda_i$, $\dot \lambda_i$ and $\ddot \lambda_i$
are found solving by iteration the system of three nonlinear equations
\begin{equation}
\bms{q'}_{\!i}^2=1 \ , \ \ \ \ \ \ \bms{q'}_{\!i} \, \bms{\cdot} \,
\bms{\dot{q}'}_{\!i}=0 \ , \ \ \ \ \ \  \bms{\dot{q}'}_{\!i}^2 +
\bms{q'}_{\!i} \, \bms{\cdot} \, \bms{\ddot{q}'}_{\!i}=0 \ . \ \ \ \ \ \ \
\end{equation}
As in the general case (8), the iteration procedure is initiated by putting
$\lambda_i = \dot \lambda_i = \ddot \lambda_i=0$ in nonlinear terms, and
unknown quantities quickly tend to the physical solutions $\lambda_i
\sim {\mit\Delta}t^3$, $\dot \lambda_i \sim {\mit\Delta}t^2$ and $\ddot
\lambda_i \sim {\mit\Delta}t$.

\vspace{8pt}

\subsection{Verlet algorithm in velocity form}

\hspace{1em}  There are the well-known group of integrators comprising Verlet
[19], leapfrog [20], velocity Verlet [21] and Beeman [22] methods. Due to
their simplicity and exceptional numerical stability they play an important
role in the classical methodology of molecular dynamics. All or some of
these methods are always described and compared in any modern textbook
[13, 14, 20, 23, 24]. However, the mentioned above approaches, being
constructed initially for the integration of Newton's equations for
translational motion, are not necessarily applicable directly to rotational
dynamics. To our knowledge, only the leapfrog method has its versions for
rotational motion [13]. The reason of such a situation is that contrary to
translational dynamics, the second time derivatives of variables, associated
with rotational degrees of freedom, may depend on their first time
derivatives. In our case the pattern is complicated additionally by the
necessity of including constraints in the equations of motion. We shall show
now how to solve these problems within the Verlet algorithm in velocity form.

Let $\{\bms{r}_i(t), \bms{\dot r}_i(t), \bms{q}_i(t), \bms{\dot q}_i(t)\}$
be a spatially-velocity configuration of the system at time $t$ and
$\{\sigma_i(t) \equiv \bms{q}_i(t)^2-1=0, \, \dot \sigma_i(t) \equiv 2
\bms{q}_i \bms{\cdot} \, \bms{\dot{q}}_i=0\}$. The translational part
$\{\bms{r}_i(t), \bms{\dot r}_i(t)\}$ of variables is considered within
the Verlet algorithm in the usual way [21, 24], whereas the rotational
variables $\{\bms{q}_i(t), \bms{\dot q}_i(t)\}$ can be evaluated as
follows. Using the principal torques $\bms{K}_i(t)$, we define angular
accelerations $\bms{\dot{\mit \Omega}}_i(t)$ and, therefore, second time
derivatives $\bms{\ddot q}_i(t)$ (4) on the basis of equations (1) for
rotational motion. Then, taking into account the constraint forces
$\bms{f}_i(t)=-2 \lambda_i(t) \bms{q}_i(t)$ yields
\begin{equation}
\bms{q}_i(t+{\mit\Delta}t) = \bms{q}_i(t) + \bms{\dot q}_i(t)
{\mit\Delta}t + \bms{\ddot q}_i(t) {\mit\Delta}t^2/2 +
\bms{f}_i(t) {\mit\Delta}t^2/2 + {\cal O}({\mit\Delta}t^3) \ . \ \ \ \ \ \
\end{equation}
The Lagrange parameters $\lambda_i$ are defined from the constraint
relations $\sigma_i(t+{\mit \Delta} t) \equiv \linebreak \bms{q}_i^2(t+
{\mit\Delta}t)-1=0$ which constitute a single quadratic equation per
molecule with the following solutions
\begin{equation}
\mathop{{\lambda_i}_1} \limits_{\ \ \ \! 2} = \frac{1}{{\mit \Delta} t^2}
\left[ 1 - \bms{\dot q}_i^2 {\mit\Delta} t^2/2 \mp \sqrt{1 -
\bms{\dot q}_i^2 {\mit\Delta} t^2 - \bms{\dot q}_i \bms{\cdot}
\bms{\ddot q}_i {\mit\Delta} t^3 - \left( \bms{\ddot q}_i^2 -
\bms{\dot q}_i^4 \right) {\mit\Delta} t^4/4} \ \right] \ , \
\end{equation}
where the time derivatives of quaternions are taken at time $t$. As can be
verified easily, only the first solution is in self-consistency with
the integration scheme. In the limit of small time steps, this solution
behaves as ${\lambda_i}_1 \to \bms{\dot q}_i \bms{\cdot} \bms{\ddot q}_i
{\mit \Delta} t / 2$, i.e., $\bms{f}_i(t) \sim {\mit \Delta} t$. Therefore,
the constraint forces contribute into the quaternion dynamics (11) terms of
order ${\mit\Delta}t^3$, i.e., the same order as numerical errors of the
used algorithm, but the rigidity of molecules is now fulfilled exactly,
i.e., $\sigma_i(t+{\mit \Delta} t)=0$.

And now we consider how to perform the second step
\begin{equation}
\dot s(t+{\mit\Delta}t) = \dot s(t) + \Big( \ddot s(t) +
\ddot s(t+{\mit\Delta}t) \Big) {\mit\Delta}t /2 +
{\cal O}({\mit\Delta}t^3) \ \ \ \ \ \ \ \ \
\end{equation}
of the velocity Verlet method, where $s$ denotes a spatial coordinate. There
are no problems to pass this step in the case of translational motion, when
$s \equiv \bms{r}_i$ and $\dot s \equiv \bms{v}_i$ is the translational
velocity. However, the difficulties immediately arise for rotational motion,
because then the second time derivative $\ddot s$ can depend explicitly not
only on the spatial coordinate $s$, but on the generalized velocity $\dot s$
as well. For example, choosing $s \equiv \bms{q}_i$, we obtain on the basis
of equations of motion (1) and (4) that $\bms{\ddot q}_i(t) \equiv
\bms{\ddot q}_i(\bms{q}_i(t), \bms{\dot q}_i(t))$. In view of (13) this
leads to a very complicated system of four nonlinear equations per molecule
with respect to four unknown components of the quaternion velocity $\bms
{\dot q}_i(t+{\mit\Delta}t)$. It is necessary to note that analogous
problems appear at attempts to apply the leapfrog, usual Verlet and
Beeman methods for rotational motion (even much more difficult in
the last two cases).

An alternative has been found in a rotational motion version [13] of the
leapfrog algorithm. It has been suggested to associate the quantity $\dot s$
with the angular momentum $\bms{l}_i = \bms{\rm A}_i^{\bvs{+}} \bms{L}_i$
of the molecule in the laboratory system of coordinates, i.e., $\dot s
\equiv \bms{l}_i$, where $\bms{L}_i = (J_1 {\mit \Omega}_1^i, J_2 {\mit
\Omega}_2^i, J_3 {\mit \Omega}_3^i) = \bms{\rm J} \bms{{\mit \Omega}}_i$
and $\bms{\rm J}$ is the diagonal matrix of principal moments of inertia.
Then the equation (13) is simplified,
\begin{equation}
\bms{l}_i(t+{\mit\Delta}t) = \bms{l}_i(t) +
\Big( \bms{k}_i(t) + \bms{k}_i(t+{\mit\Delta}t) \Big) {\mit\Delta}t /2 +
{\cal O}({\mit\Delta}t^3) \ \ \ \ \ \ \ \ \ \
\end{equation}
and, therefore, $\bms{l}_i(t+{\mit\Delta}t)$ are easily evaluated using
the torques $\bms{k}_i(t+{\mit\Delta}t)$ in the new spatial configuration
$\{\bms{q}_i(t+{\mit\Delta}t)\}$. At the same time, new values for principal
angular and quaternion velocities are obtained (when they are needed) using
the relations $\bms{\mit \Omega}_i(t+{\mit\Delta}t)=\bms{\rm J}^{-1}
\bms{\rm A}_i(t+{\mit\Delta}t) \bms{l}_i(t+{\mit\Delta}t)$ and
$\bms{\dot q}_i(t+{\mit\Delta}t)=\frac12 \bms{\rm Q} (\bms{q}_i(t+
{\mit\Delta}t)) \bms{\Omega}_i(t+{\mit\Delta}t)$.

Finally, we consider the third version of the velocity Verlet method for
rotational motion. The idea consists in using angular velocities as
independent parameters for describing the sate of the system in phase
space. Then choosing $\dot s \equiv \bms{\mit \Omega}_i$ and taking into
account Euler equations (1), we obtain from (13) the following result
\begin{eqnarray}
{{\mit \Omega}_\alpha^i}^{(n)}(t+{\mit\Delta}t) =
{\mit \Omega}_\alpha^i(t) + \frac{{\mit\Delta}t}{2 J_\alpha } \bigg[
K_\alpha^i(t) + K_\alpha^i(t+{\mit\Delta}t) \hspace{4.6cm}
\nonumber \\ [-5mm] \\ \hspace{1.8cm}
+ \Big(J_\beta-J_\gamma\Big) \Big({\mit \Omega}_\beta^i(t)
{\mit \Omega}_\gamma^i(t) +{{\mit \Omega}_\beta^i}^{(n-1)}(t+{\mit\Delta}t)
{{\mit \Omega}_\gamma^i}^{(n-1)}(t+{\mit\Delta}t) \Big) \bigg] \ .
\nonumber
\end{eqnarray}
Unless $J_1=J_2=J_3$, the equations (15) are, in fact, the system of
three quadratic equations per molecule with respect to the three unknowns
${{\mit \Omega}_\alpha^i}(t+{\mit\Delta}t)$. The system (15) is relatively
simple and can be solved by iteration $(n=1,2,\ldots)$ with ${{\mit
\Omega}_\alpha^i}^{(0)}(t+{\mit\Delta}t) = {\mit \Omega}_\alpha^i(t)$ as
an approximation of zero order. A few iterations is sufficient for actual
time steps to find the desired solutions with a great precision.

From a mathematical point of view, all the three representations $\dot s
\equiv \bms{\dot q}_i, \bms{l}_i$ or $\bms{\mit \Omega}_i$ are completely
equivalent, because the knowledge of an arbitrary quantity from the set
$(\bms{\dot q}_i, \bms{l}_i, \bms{\mit \Omega}_i)$ allows us to determine
uniquely the rest of two ones. In the case of numerical integration the
pattern is qualitatively different, because the investigated quantities
are evaluated approximately. The choice $\dot s \equiv \bms{\dot q}_i$
can not be recommended for calculations due to its complexity. Moreover,
it has yet a major disadvantage that the equality $\bms{q}_i \bms{\cdot}
\, \bms{\dot{q}}_i=0$ is broken at time $t+{\mit\Delta}t$, whereas this
equality remains in law by construction in other two cases.

The case $\dot s \equiv \bms{l}_i$ is the most attractive in view of the
avoidance of nonlinear equations. Computations show (see Sec.~4), however,
that the best numerical stability with respect to the total energy
conservation exhibits the third version, when $\dot s \equiv \bms{\mit
\Omega}_i$. The reason of this can be found, taking into account that a
kinetic part, $\frac12 \sum_{i=1}^N (J_1 ({\mit \Omega}_1^i)^2+J_2 ({\mit
\Omega}_2^i)^2+J_3 ({\mit \Omega}_3^i)^2)$, of the total energy is
calculated directly from principal angular velocities. At the same time,
to evaluate angular velocities within the previous approach the additional
transformations $\bms{\mit \Omega}_i=\bms{\rm J}^{-1} \bms{\rm A}_i
\bms{l}_i$ with approximately computed matrices $\bms{\rm A}_i=\bms{\rm A}
(\bms{q}_i)$ and angular momenta $\bms{l}_i$ are necessary. They contribute
an additional portion into accumulated errors at the calculations of the
total energy.

\vspace{12pt}

\section{Numerical results and discussion}

\hspace{1em}  We now test our integration approach on the basis of MD
simulations for a TIP4P model [25] of water. The simulations were
performed in the microcanonical ensemble at a density of 1 g/cm$^3$
and at a temperature of 298 K. We considered a system of $N=256$ molecules
in the reaction field geometry [26]. All runs were started from an identical
well equilibrated configuration. Numerical stability was identified in
terms of the relative fluctuations ${\scr E}_t=\sqrt{\left< (E-\left<E
\right>_t)^2 \right>_t}\Big/\left<E\right>_t$ of the total energy of the
system during time $t$.

Samples of the function ${\scr E}_t$ for the fourth-order Gear algorithm
are presented in fig.~1. It can be seen easily from the figure that within
the modified version, where quaternion-constraint transformations (9) are
applied, the conservation of the total energy is improved considerably in
comparison with that obtained within the original version of the algorithm,
when no additional corrections are used. The system of nonlinear equations
(10) were solved with the relative iteration precision of $10^{-12}$ and
the number of iterations never exceeded 5. At the same time, applying
the rescaling scheme leads even to worse results than those given by the
original version. Despite the fact that the Gear algorithm integrates the
equations of motion very well at ${\mit \Delta}t \le 1$ fs, it has a very
small region of stability with respect to time steps [16] and can not be
used for ${\mit \Delta}t \ge 1.5$ fs (see fig.~1d). Moreover, the scheme
predictor-corrector-corrector, which was chosen by us at ${\mit \Delta}t
\le 1$ fs to provide an optimal performance, takes computational time per
step twice larger than the Verlet integrator (in the case ${\mit \Delta}t
= 1.5$ fs, three corrector steps were used).

As the atomic-constraint technique [15, 16] is intensively exploited and
its advantages with respect to numerical stability are generally recognized,
we have made comparative test carrying out explicit MD runs using this
method as well as the angular-velocity Verlet algorithm (15) within the
quaternion-constraint dynamics (11). The usual value of the time step
in simulating such a system is ${\mit \Delta}t=2$ fs [27]. These two
approaches required almost the same computer time per step (96\% being
spent to evaluate pair interactions). The corresponding functions
${\scr E}_t$ are shown in fig.~2 at four fixed values ${\mit \Delta}t=$
1, 2, 3 and 4 fs. For the purpose of comparison the results obtained
within the angular-momentum version (14) are also included in this
figure (insets (a), (b)). The rescaling scheme (instead of the
quaternion-constraint dynamics) within the angular-velocity Verlet
version was considered as well.

At the smallest time step (fig.~2a), all the approaches exhibit similar
equivalence in the energy conservation. The time step ${\mit \Delta}t=1$ fs
is too small, however, and impractical in simulations because it requires
too much computation time to cover the sufficient phase space. For larger
time steps (fig.~2b-d), the total energy fluctuations increase drastically
with increasing the length of the runs within the usual quaternion rescaling
scheme, whereas the atomic- and quaternion-constraint methods conserve the
energy approximately with the same accuracy. As far as the rescaling of
quaternions has been chosen, it must be applied after each time step to
achieve an optimal performance within the Verlet integrator. For example,
if no corrections are performed, the rigidity of molecules is broken
catastrophically (see the dotted curve in fig.~2b). As we can see, the
modified quaternion approach always leads to improved results. Finally,
we show in fig.~2a,b that the angular-momentum version of the Verlet
algorithm is much more unstable and can be used at small (${\mit \Delta}t
\le 1$ fs) time steps only. For these reasons, taking into account also
comments on the Gear algorithm, the crude renormalization procedure is
generally not recommended and preference must be given to the modified
quaternion approach within the angular-velocity Verlet integrator. Quite
a few iterations (the mean number of iterations per molecule varied
from 2 to 4 at ${\mit \Delta}t =$ 1 -- 4 fs) was sufficient to obtain
solutions to the system of nonlinear equations (15) with the relative
iteration precision of $10^{-12}$. This required a negligible small
computation time additionally to the total time.

The calculations have shown that the same level ${\scr E}=0.025 \%$ of
energy conservation can be provided by the time steps of 2.1, 3.7 and 4.0 fs
within the quaternion rescaling, atomic- and quaternion-constraint schemes.
Therefore, the last two approaches allow a time step approximately twice
larger than the quaternion rescaling method. A reason of this gain in
time can be explained by the fact that the rescaling of quaternions is
an artificial procedure. It involves an unpredictable discrepancy in the
calculation of trajectories of atoms at each step of the integration process
and leads to significant deviations of the total energy. At the same time,
the atomic- and quaternion(molecular)-constraint techniques provide the
rigidity of molecules in a more natural way.

\vspace{12pt}

\section{Conclusion}

\hspace{1em}  An alternative scheme to overcome the difficulties
in simulations of rigid polyatomics has been proposed. The scheme
uses the constraint formalism for treating quaternion dynamics. As
a result, the rigidity problem has been rigorously resolved, using
quaternion-constraint forces. Although this introduces some extra
transformations, but presents no numerical difficulties. In a particular
case of the velocity Verlet algorithm, the constraint version of the
quaternion approach allows one to perfectly fulfil the rigidity of
molecules at each step of the trajectory without any additional efforts
and loss of precision. Avoidance of the necessity to solve complex
nonlinear equations for maintaining molecular rigidity should be a
benefit of the presented approach with respect to the atomic-constraint
dynamics.

It has been corroborated by actual MD simulations that the quaternion
rescaling method is much less efficient than the atomic- and
quaternion-constraint techniques. The last both schemes seem to be
comparable in efficiency. The advantage of the modified quaternion
approach is that it improves the energy conservation considerably
at a little computational cost.

\vspace{18pt}

{\bf Acknowledgements.} The author would like to acknowledge financial
support of the President of Ukraine.

\newpage
\thispagestyle{empty}

\vspace*{1cm}

\begin{center}
{\large Figure captions}
\end{center}

\vspace{12pt}

{\bf Fig.~1.}~The relative total energy fluctuations as functions of
the length of the simulations performed within three versions of the
Gear algorithm at four fixed time steps. Note, that all three curves
are indistinguishable in {\bf (a)}.

\vspace{12pt}

{\bf Fig.~2.}~The relative total energy fluctuations as functions of
the length of the simulations performed within the atomic-constraint
technique and various versions of the velocity Verlet algorithm at four
fixed time steps. The dotted curve in {\bf (b)} corresponds to the usual
quaternion approach without any additional corrections. Note, that three
of four curves are indistinguishable in {\bf (a)}.

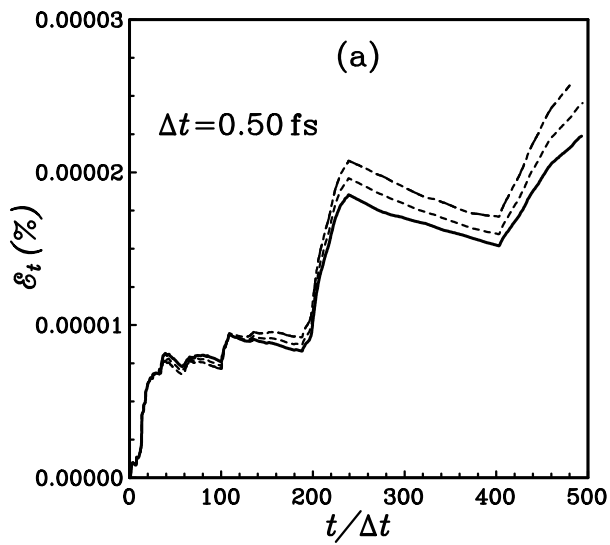
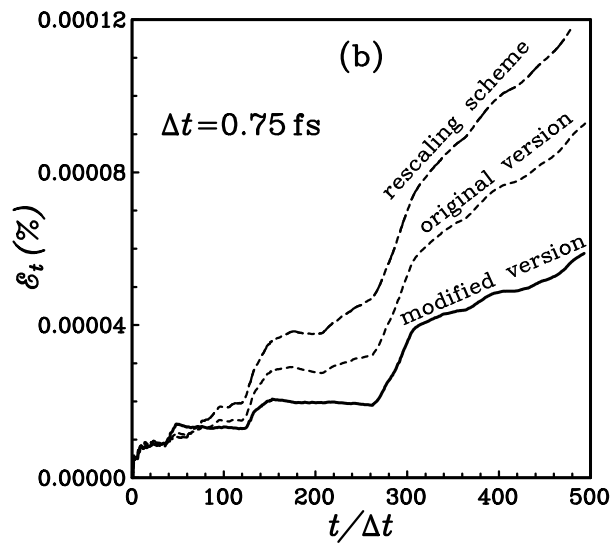
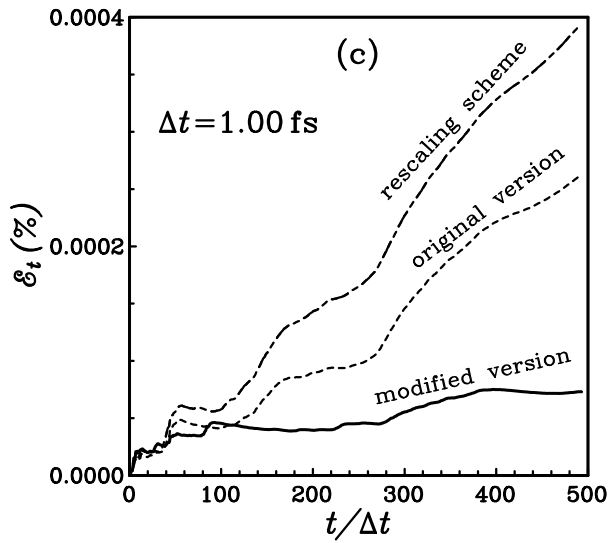
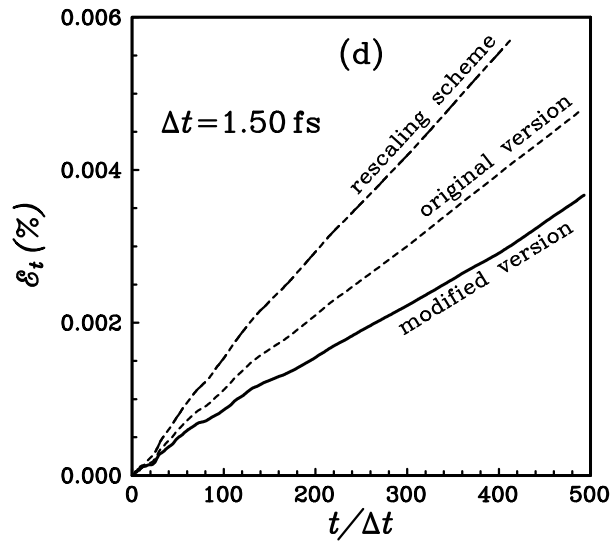
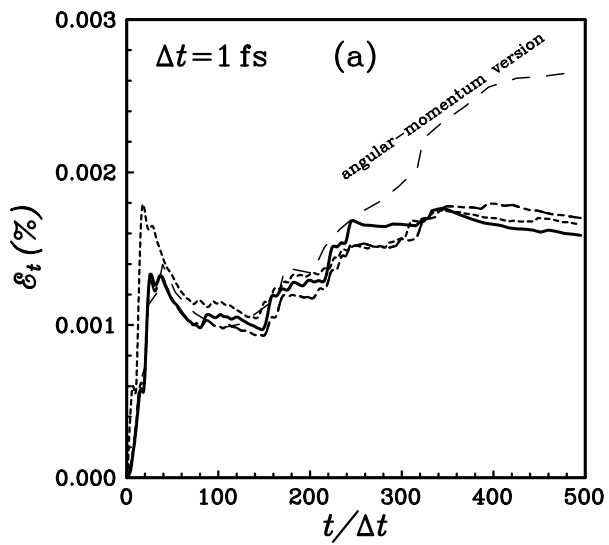
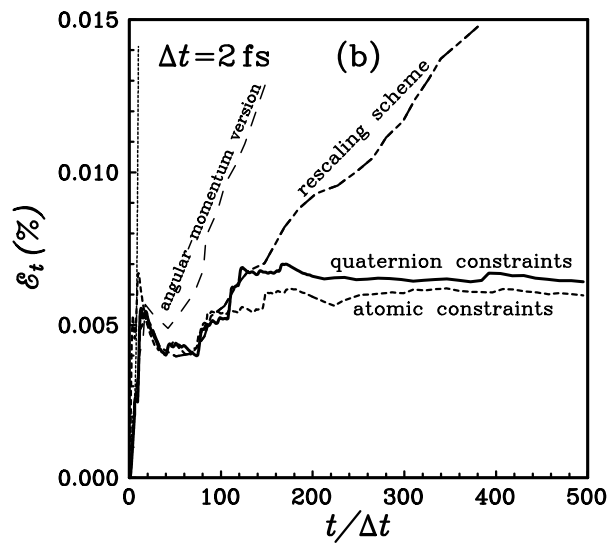
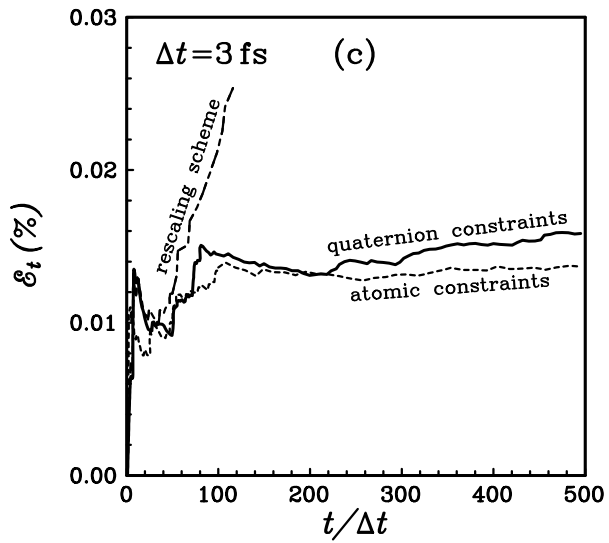
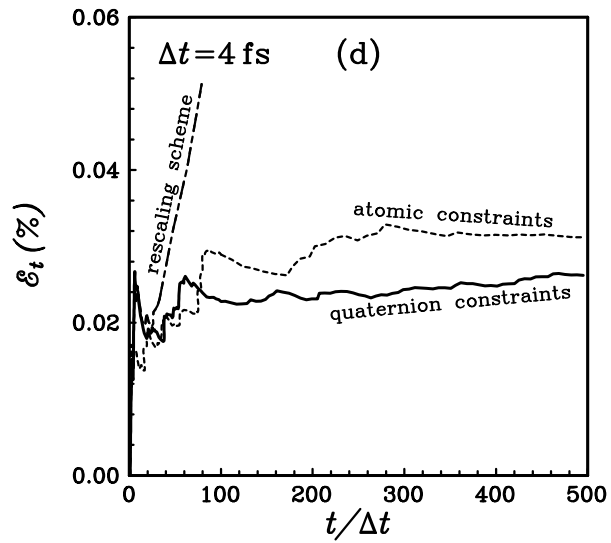
\end{document}